# Proposal for Cherenkov Time of Flight Technique with Picosecond Resolution

S. Majewski[1], A. Margaryan[2], L. Tang[1]

[1] *Thomas Jefferson National Accelerator Facility, Newport News, VA 23606, USA*
[2] *Yerevan Physics Institute, 375036 Yerevan, Armenia*

A new particle identification device for Jlab 12 GeV program is proposed. It is based on the measurement of time information obtained by means of a new photon detector and time measuring concept. The expected time measurement precision for the Cherenkov time-of-flight detector is about or less than 10 picosecond for Cherenkov radiators with lengths less than 50 cm.

## 1. Introduction

Time measurements are used in high energy particle and nuclear physics experiments to identify particle types, and to distinguish background from real events (see e.g. [1]). In some cases precise time-of-flight (TOF) measurements can provide a better measurement of the particle energy than the magnetic spectrometers. The time precision limit for the current systems consisting of particle detectors based on vacuum photomultiplier tube -PMTs or hybrid photon detector -HPDs, timing discriminators and time to digital converters is about 100 ps (FWHM).

However, it is well known that timing systems based on RF fields can provide ps precision. Streak cameras, based on similar principles, are used routinely to obtain picosecond time resolution of the radiation from pulsed sources [2-4]. The best temporal resolution is achieved with a single shot mode. Using continuous circular image scanning with $10^{-10}$s period, a time resolution of $5 \times 10^{-13}$ s for single photoelectrons and $5 \times 10^{-12}$ s for pulses consisting of ~17 secondary electrons, induced by heavy ions, was achieved in early seventies [5]. With a streak camera operating in the repetitive mode known as "synchroscan", typically, a temporal resolution of 2 ps (FWHM) can be reached for a long time exposure (more than one hour) by means of proper calibration [6]. But, the streak cameras are expensive, their operation is complicated and they did not find wide application in the past in the elementary particle and nuclear physics experiments.

The basic principle of a RF timing technique or a streak camera is the conversion of the information in the time domain to a spatial domain by means of ultra-high frequency RF deflector. Recent R&D work (ISTC Project A-372, [7]) has resulted in an effective 500 MHz RF circular sweep deflector for keV energy electrons. The sensitivity of this new compact RF deflector is about V/mm and is an order of magnitude higher than the sensitivity of the RF deflectors used previously. An essential progress in this field can be reached by a combination of a new RF deflector created at YerPhI with new technologies for position sensors, e.g. position sensitive electron multipliers from Burle Industries Inc., USA [8] and of a new concept for a compact and easy-to-use photon detector and timing concept with picosecond resolution proposed recently in [9].

In this paper we consider a Cherenkov time-of-flight detector based on the new photon detector and the time measuring concept.

## 2. Picosecond Photon Detector

The principle of the picosecond photon detector is to couple in vacuum the position sensitive technique to an input window with photocathode evaporated on its inner surface (Fig.1). The light photon incident on the photocathode generates a photoelectron, which is subsequently accelerated up to several keV, focused, and transported to the RF resonator deflector. Passing through the RF deflector, the electrons change their direction as a function of the relative phase of electromagnetic field oscillations in the RF cavity, thereby transforming time information into deflection angle. In this project, we propose to use the oscilloscopic method of RF timing, which will result in a circular pattern on the detector plane. The coordinates of the secondary electrons are measured with the help of a position sensitive detector located at some distance behind the RF deflecting system. The architecture of the technique is very similar to the design of the HPDs [10-12]. In contrast to ordinary PMTs, where a photoelectron initiates a multi-step amplification process in a dynode system, in the HPDs a photoelectron undergoes single-step acceleration from the photocathode towards the position sensitive detector where electron multiplication process happens in the solid state anode target material.

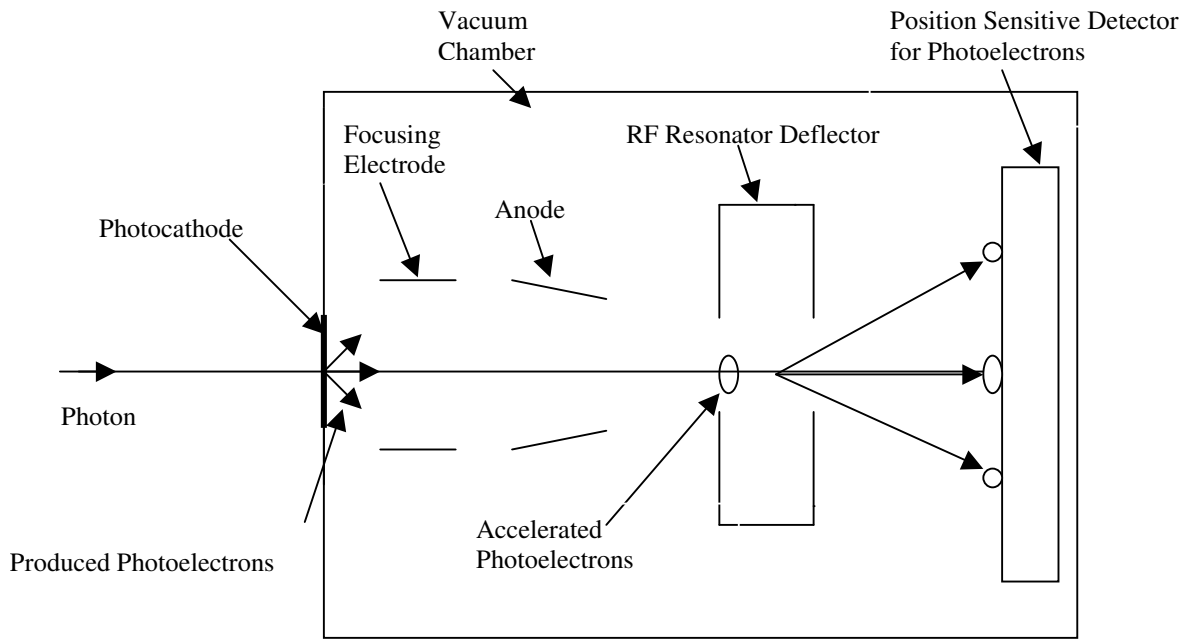

FIG. 1: A schematic drawing of the picosecond photon detector.

Recently at YerPhI, we have successfully employed a dedicated RF deflecting system at a frequency of about f = 500 MHz and obtained a circular scan of a 2.5 keV electron beam on the phosphor screen (see Fig. 2). This compact, simple and effective RF deflecting system is very suitable for high frequency RF analyzing of few-keV electrons.

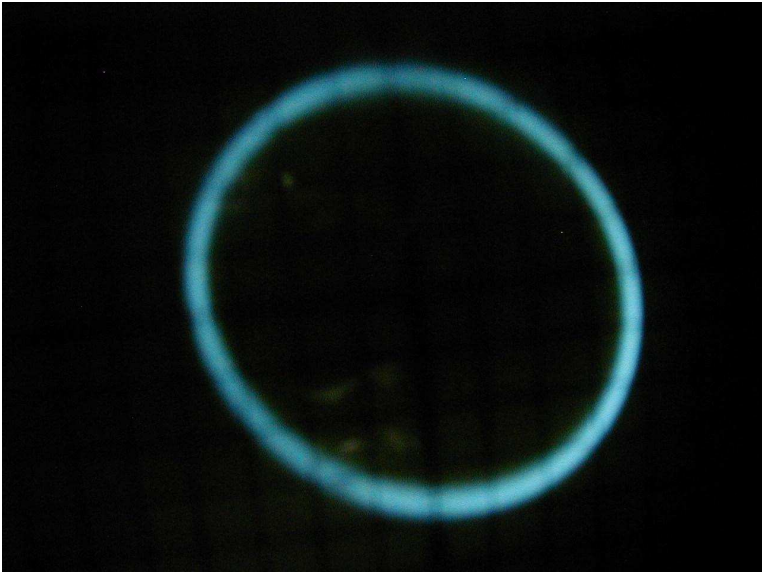

FIG. 2: Image of a circularly scanned 2.5 keV electron beam on the phosphor screen.

The time resolution of the system presented in Fig.1 depends on several factors [13, 14]:

## 2.1 Physical Time Resolution of Photocathode

General quantum-mechanical considerations show that the inherent time of the photoelectron emission should be much shorter than $10^{-14}$s. However, a delay and time spread $\Delta\tau_p$, of the electron signal can be caused by the finite thickness of the photocathode $\Delta l$ and the energy spread of the secondary electrons $\Delta\varepsilon$. For $\Delta l = 10$nm and $\Delta\varepsilon = 1$ eV we obtain $\Delta\tau_p \approx 10^{-13}$s.

## 2.2 Physical Time Resolution of Electron Tube

The minimal physical time dispersion of the electron optics is determined by chromatic aberration due to the electron initial energy spread $\Delta\varepsilon$. The time spread, $\Delta\tau_t$, due to this effect in the case of a uniform accelerating electric field E in V/cm near the photocathode plane is $\Delta\tau_t = 2.34 \times 10^{-8}$ $(\Delta\varepsilon)^{1/2}$/E s. For $\Delta\varepsilon = 1$ eV and E = 10 kV/cm we obtain $\Delta\tau_t \approx 2$ ps.

## 2.3 Technical Time Resolution of Electron Tube

Electron tube is a device with precise electron focusing with effective electron transit time equalization. The time precision limit for the whole system consisting of photo-cathode, accelerating, focusing and transporting parts in the carefully designed system is estimated to be on the order of 0.1 ps (FWHM) for point like photocathode [15] and 10 ps (FWHM) for active photocathode diameter of 4 cm [16] or even larger [17].

## 2.4 Technical Time Resolution of RF Deflector

By definition, the technical time resolution is: $\Delta \tau_d = d / v$, where $d$ is the size of the electron beam spot or of the position resolution of the secondary electron detector if the electron beam spot is smaller, while $v$ is the scanning speed: $v = 2\pi R / T$. Here T is the period of rotation of the field, R is the radius of the circular sweep on the position sensitive detector. For example, if $T = 2 \times 10^{-9}$ s (f = 500 MHz), R = 2 cm, and d = 0.5 mm, $v \geq 0.5 \times 10^{10}$ cm/sec and we have $\Delta \tau_d \leq 10 \times 10^{-12}$ s.

## 3. Position Sensitive Detector

The position sensitive detectors for the RF timing technique should have the following properties:
- High detection efficiency for 2.5-10 keV single electrons;
- Position sensitivity, for example, $\sigma_x \leq$ of 0.5 mm;
- High rate capability.

These three requirements are to a different degree realized by several types of detectors. They include vacuum- based devices such as the Multi-anode PMT [18,19], Micro Channel Plate (MCP) [20] as well as arrays of Si PIN [21] and Avalanche Photodiodes (APD) [22] or APD working in a Geiger mode (SiPM) [23].

However, the best approach is the development of a dedicated multi-anode PMT with circular anode structure.

Position determination can be performed in two basic architectures:
1) Direct readout: array of small pixels, with one readout channel per pixel, such as available with avalanche Si diodes. The position resolution in this case is about or better than the size of readout cell.
2) Interpolating readout: position sensor is designed in such a way that the measurement of several signals (amplitudes and/or times) on neighboring electrodes defines event position. The position resolution limit for both cases is $\Delta x / x \sim 10^{-3}$ [24].

From the temporal resolution budget discussed above one can conclude that the expected parameters of the technique are:
a) Internal time resolution for each photo-electron of about 2 ps;
b) Technical time resolution in the range of $(2-10) \times 10^{-12}$ s for f = 500 MHz RF deflector;
c) Absolute calibration of the system better than $10^{-12}$ s is possible, which is an order of magnitude better than can be provided by a regular timing technique.

After the analysis of the available technical solutions for electron multipliers in our opinion the best solution is the development of a dedicated PMT based on MCP with circular anode structure and with both readout architectures for position determination with relative precision close to $10^{-3}$. This technique also provides an optical waveform digitization with a bandwidth of >10 GHz and can be used, e.g., for digitization of Cherenkov radiation flash from elementary particles.

## 4. Cherenkov TOF Detector Based on Picosecond Photon Technique

The time resolution of the TOF counter using plastic scintillators is inherently limited by the following effects besides the transit-time spread (TTS) of the phototube:

i)      a finite decay time of photon emission;

ii)     a different photon propagation-length or propagation-time in the scintillators to a phototube depending on the photon emission angle.

Thus the conventional TOF counter measures the arrival time of the earliest photons out of many, whereby the remaining large amount of photons are not effectively used for the time measurement.

By using Cherenkov radiation rather than the scintillation photons, two substantial and undesirable effects can be eliminated. While a decay constant of fast counting scintillator is typically ~2 ns, the flash duration of the Cherenkov radiation is ≤ 1ps [25]. While the propagation paths of scintillation photons are not unique due to uniform emission over $4\pi$ solid angle, they are unique for the Cherenkov radiation. The Cherenkov photon emission angle, $\theta_c$, is uniquely determined by the particle velocity $\beta$. Depending on the particle, its momentum range, and experimental condition, there are various ways to do this using liquid, gas, or solid radiators with different type of photon detectors.

We will consider Cherenkov TOF detector in which the good time resolution is crucial. Similar to the DIRC, Detection of Internally Reflected Cherenkov light [26], the Cherenkov TOF couples DIRC bar with a picosecond photon detector that times the photoelectrons seen at the end of the bar. In so doing, a DIRC utilizes the optical material of the radiator in two ways, simultaneously; first, as a Cherenkov radiator, and second, as a light pipe for the Cherenkov light trapped in the radiator by total internal reflection. The principal structure of the DIRC in the (x,y,z) coordinate system (bar frame) is illustrated in Fig.3.

Each radiator is a long, thin bar with a rectangular cross section of transverse dimensions ($t_x, t_y$). When a charged particle passes through the radiator bar, Cherenkov photons are emitted in a conical direction defined by the Cherenkov angle $\theta_c$, where $\cos\theta_c = 1/n\beta$, n is the refractive index. The source length of the emitting region is the particle trajectory length in the radiation material. The angles, positions and momentum of the incident particle are provided by other detectors.

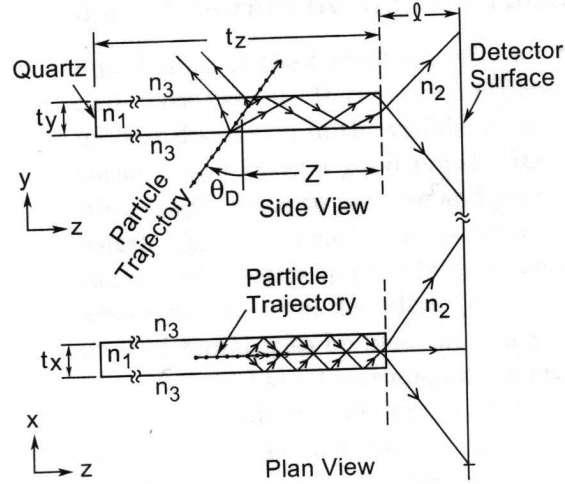

FIG. 3: Schematic of a radiator bar of a DIRC counter illustrating the principle of the device. The particle trajectory is shown as a line connected by dots; representative trajectories of Cherenkov photons are shown by lines with arrows.

In the three dimension DIRC concept, the high reflection coefficients inherent in the total internal reflection process, and the fact that the magnitudes of angles are conserved during reflection from a flat surface allow the photons of the ring image to be transported to a detector outside the path of the radiating particle, where they may be imaged. Denoting the polar and azimuthal angles of an incident charged particles as $\theta_p$ and $\phi_p$, respectively, the directional cosines of the photon emission in the bar frame can be written as:

$$q_x = -q_{x'}\cos\theta_p\cos\phi_p + q_{y'}\sin\phi_p + q_{z'}\sin\theta_p\cos\phi_p,$$
$$q_y = -q_{x'}\cos\theta_p\cos\phi_p - q_{y'}\cos\phi_p + q_{z'}\sin\theta_p\sin\phi_p,$$
$$q_z = q_{x'}\sin\theta_p + q_{z'}\cos\theta_p,$$

where $q_i$'s$(i = x',y',z')$ are the directional components of the photon emission in the frame where the particle moves along the $z'$-axis:

$$q_{x'} = \sin\theta_c\cos\phi_c,$$
$$q_{y'} = \sin\theta_c\sin\phi_c,$$
$$q_{z'} = \cos\theta_c.$$

The horizontal and vertical photon angles at the bar end are then given as

$$\Phi = \arctan (q_x/q_z),$$
$$\Theta = \arctan (q_y/q_z).$$

The photon propagates a length (L$_p$) in a Time-Of-Propagation -TOP (t$_p$), down a bar length of (L) as is given by

$$t_p = \frac{L_p n(\lambda)}{c} = \frac{Ln(\lambda)}{cq_z} \, ,$$

where $n(\lambda)$ is the refractive index of the radiator at wavelength $\lambda$, c is the light velocity in the vacuum, and $q_z$ is the directional z-component of the photon emission.

The azimuthal angle $\phi_c$ of Cherenkov photons is distributed uniformly over $2\pi$. When one fixes $\phi_c$, the individual three-directional components $-q_i$ are uniquely related to $\theta_c$. Therefore, a measurement of any two directional components or any two combinations of them provides information about $\theta_c$. In a DIRC concept the two parameters ($\Theta, \Phi$) and time measurement for extracting $\theta_c$ are used.

The DIRC prototypes have been constructed and tested over the past. The first large-scale DIRC detector designed for physics is now running in the BABAR detector at PEP-II [27].

A number of DIRC devices have been proposed that use less than three dimensions. For example, a 1-D device called the Cherenkov Correlated Timing technique - CCT that couples DIRC bars with a non-imaging detection system that times the first photoelectron(s) seen at the bar end was proposed [28]. A prototype has been constructed and tested by Kichimi et al.[29]. The technique uses the correlation between photon path-length and Cherenkov production angle to infer this angle by measuring the time taken for the totally reflected Cherenkov photons to "bounce" to the end of the radiator. This simple principle can be illustrated by the 2-dimensional example shown in Fig. 4. Fig. 4 also demonstrates the difference in flight path for light emitted by $\pi$'s (thick line) and K's (thin line). Due to this difference the TOP is inversely proportional to z (quartz-axis direction)-component of the light-velocity, which produces TOP differences of, for instance, about 100 ps or more for normal incident 4 GeV/c $\pi$ and K at 2 m long propagation.

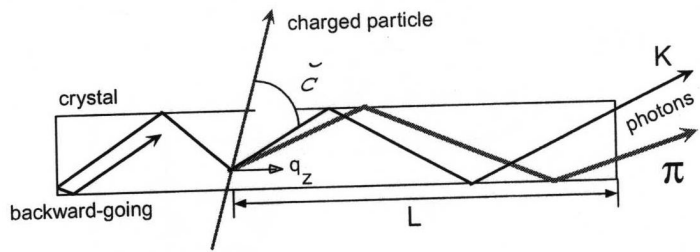

FIG. 4: Schematic of the Cherenkov correlated timing technique.

Akatsu et al. have proposed and tested [30] a 2-D readout DIRC , one timing dimension $-t_p$ and one space dimension $-\Phi$, called them the Time Of Propagation -TOP counter.

The DIRC counters have confirmed that the Cherenkov radiation images are well transported by internal total reflection, and the CCT counter exhibited that a time measurement of such photons is feasible, while the TOP detector are based on the concept of a 2-D readout DIRC, one timing dimension and one space.

We will consider TOF technique based on the measurement of the time of propagation of Cherenkov photons, similar to the one dimension DIRC- CCT counters, but detection of Cherenkov photons will be carried out by means of picosecond photon detector.

## 4.1 Simulation Study

Based on the basic concepts considered in the preceding section, we have carried out a simulation study of the proposed technique. In the modeling of the detector the parameters of the radiator are assumed as described in [28, 29]. There are several dominant contributions to the spread of the TOP of the Cherenkov photons in the radiator in addition to the time spread of photon detector:

1) The time spread of Cherenkov radiation along the particle trajectory over the thickness of quarz bar: Cherenkov photons are emitted uniformly along the path of particle passage through the radiator.

2) The transit time spread of Cherenkov photons due to different trajectories: trajectories of the individual photons determined according to $\theta_c$ and $\Phi_c$ of individual Cherenkov photons.

3) The chromatic effect of Cherenkov light: for the numerical calculations we take $n = 1.47 \pm 0.008$. The Gaussian distribution for n has been used.

4) The timing accuracy of the Photon detector: for the timing accuracy of the Photon detector we take $\sigma = 10$ ps.

The expected total number of photoelectrons- $N_{pe}$ detected in the device is: $N_{pe} = N_0 \times \Delta l \times (\Delta \Phi_c / 360)$, where $N_0 = 155$ pe/cm [29] is the Cherenkov quality factor, $\Delta l$ is the thickness of the radiator, $\Delta \Phi_c$-is the azimuthal angular interval of the detected Cherenkov photons. The size of the radiator bar is 20 mm-thick (in y), 20 mm-wide (in x), and length is in the range 10-200 cm (in z). One end of the Cherenkov radiator is connected to the photon detector through light pipe, the special design of which allows to select the azimuthal angular interval - $\Delta \Phi_c$, and the opposite end assumed blackened to avoid reflection. So we are considering only forward - FW going photons.

In order to understand the TOP behavior and to evaluate the size of the above effects to TOP and the PID power, a detail calculation has been performed. It has been demonstrated that the results of previous DIRC simulations [30] as well as of the experimental studies [29] are reproduced by our approach with precision better than 20%. Some results are displayed in the Fig.5-Fig.7 and tabulated in Table 1.

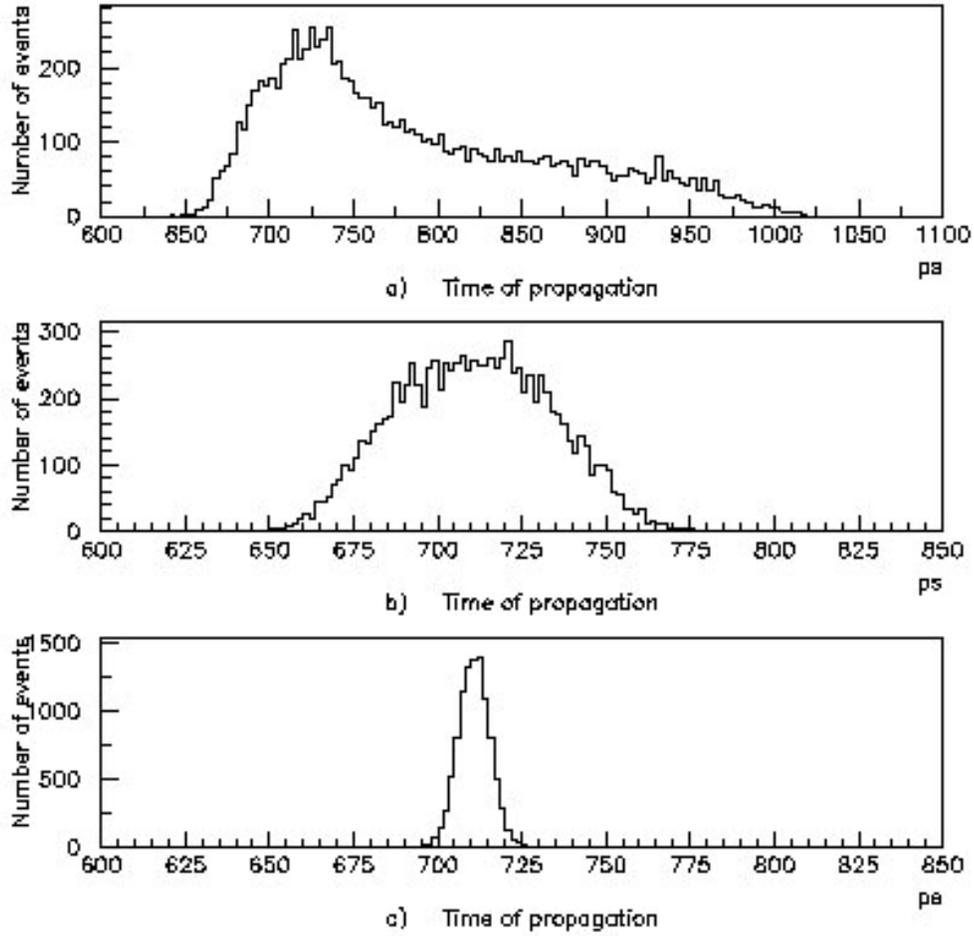

FIG. 5: Time of propagation distribution of single and FW going photons for tracks of p = 2 GeV/c pions, $\theta_{inc} = 90°$, with $|\Phi_c| \leq 45°$ (a), $|\Phi_c| \leq 15°$ (b), and mean time distribution of all detected photons (~25) for the case with $|\Phi_c| \leq 15$ (c). The radiator length is L = 10 cm and thickness $\Delta l = 2.0$ cm. For the Cherenkov quality factor we use $N_0 = 155$ pe/cm [29]. Number of events for each case is equal 10000.

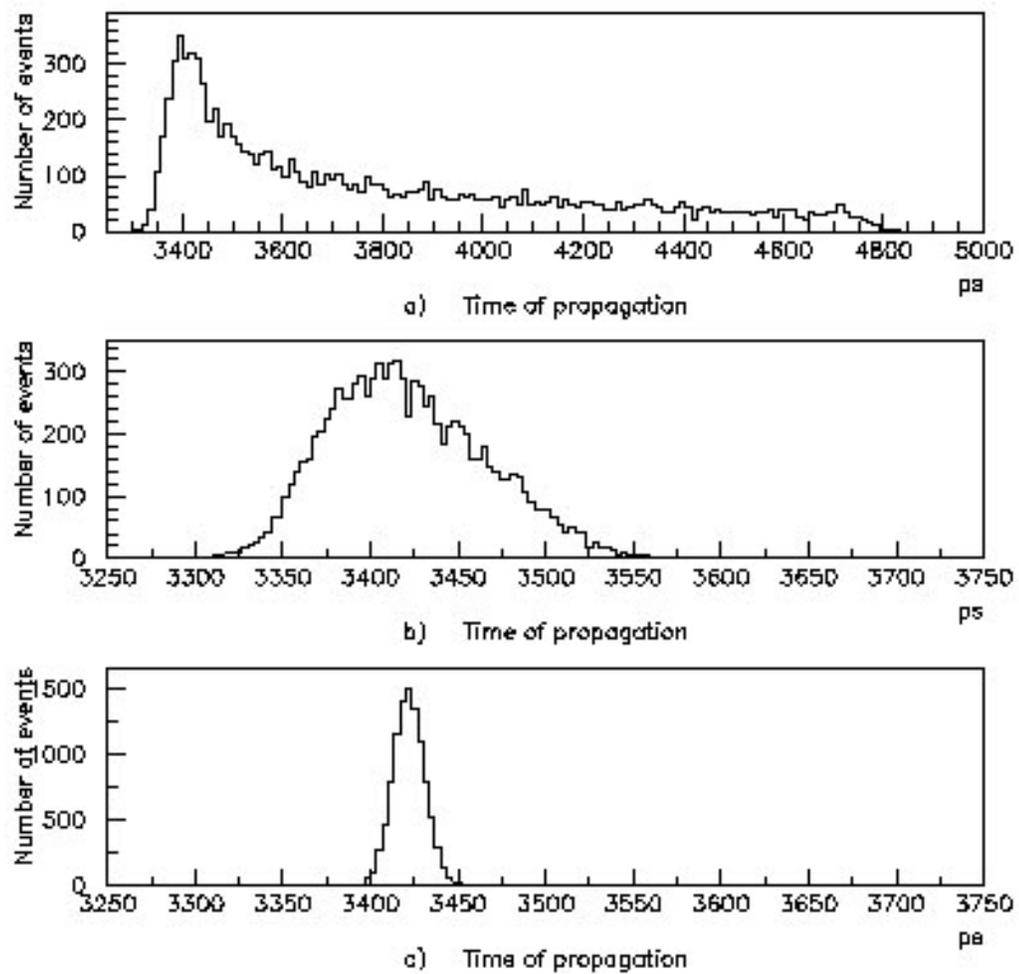

FIG. 6: The same as in Fig. 5 but with radiator length L = 50 cm.

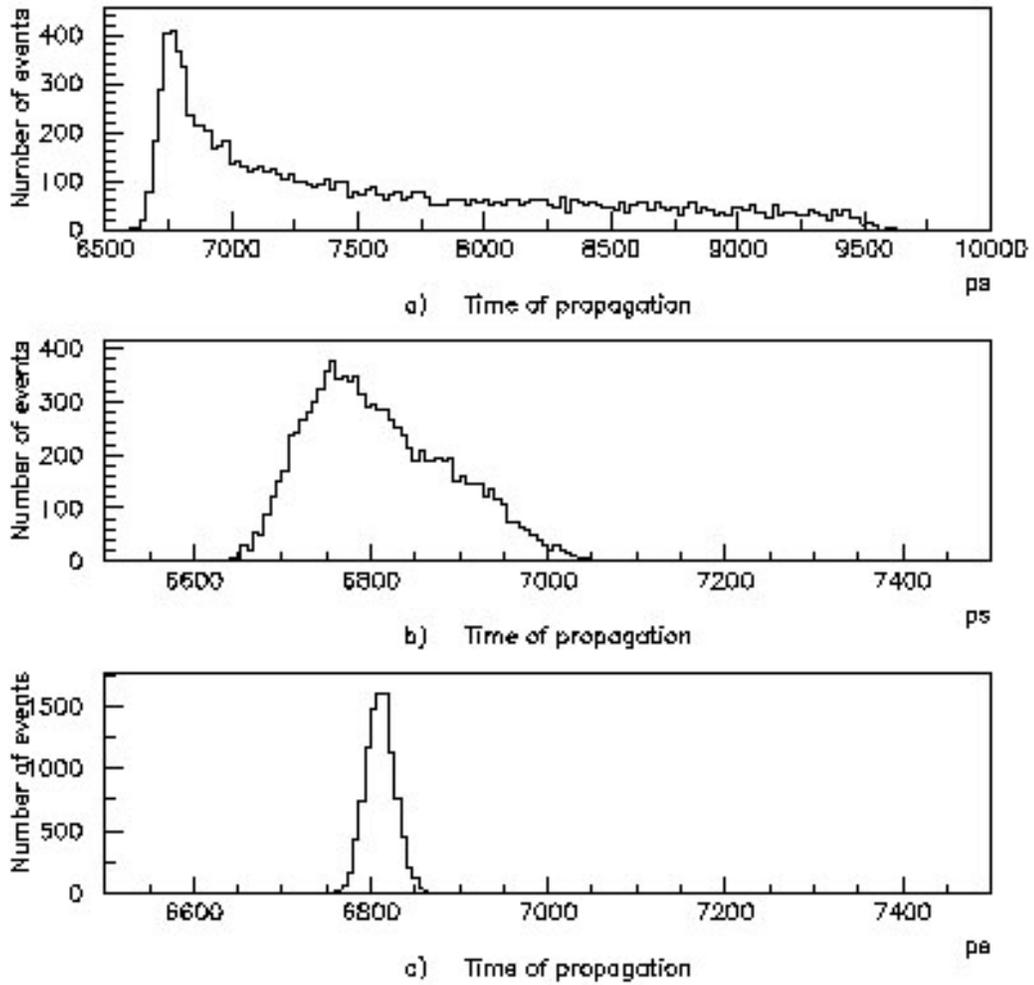

FIG. 7: The same as in Fig. 5 but with radiator length L = 100 cm.

Table 1: Average time of propagation and its spread of forward going photons for tracks of p = 2000 MeV/c $\pi$ and K with $\theta_{inc}$=90°, $|\Phi_c| \leq 15°$ and for different length L.

| L (cm) | 20 | | 40 | | 100 | | 200 | |
|---|---|---|---|---|---|---|---|---|
| Particle | $\pi$ | K | $\pi$ | K | $\pi$ | K | $\pi$ | K |
| TOP (ps) | 1388 | 1423 | 2744 | 2812 | 6811 | 6982 | 13589 | 13930 |
| $\Delta$(TOP) (ps) | 5.8 | 6.0 | 9.0 | 9.3 | 20.1 | 21 | 40 | 42 |
| TOP(K) - TOP($\pi$) (ps) | 35 | | 68 | | 171 | | 341 | |

The TOP distributions of 10000 events with 50% $\pi$ and 50% K tracks of $\theta_{inc}$=90°, $|\Phi_c| \leq 15°$, L = 100cm and for p = 1.5, 2.0 and 3.0 GeV/c momentum are displayed in

Fig.8. The number of σ separation $N_\sigma$ between π/K of p = 2 GeV/c momentum and with L = 100 cm is about 8.5. It is interesting to compare with results of the 3D DIRC. For a particle of momentum p well above threshold entering a radiator with index of refraction n, the number of σ separation $N_\sigma$ between particles of mass $m_1$ and $m_2$ in the 3D DIRC is approximately [26]:

$$N_\sigma \approx \frac{\left|m^2{}_1 - m^2{}_2\right|}{2\,p^2\sigma[\theta_{c\,(tot)}]\sqrt{n^2-1}}.$$

For the $N_\sigma = 8.5$ separation, the $\sigma[\theta_c(tot)]$ must be about 3 mrad, for n = 1.47 radiator. In the practical 3D DIRC counters, the attainable angular resolution $\sigma[\theta_c(tot)]$ varies between 0.5 and 5 mrad [26].

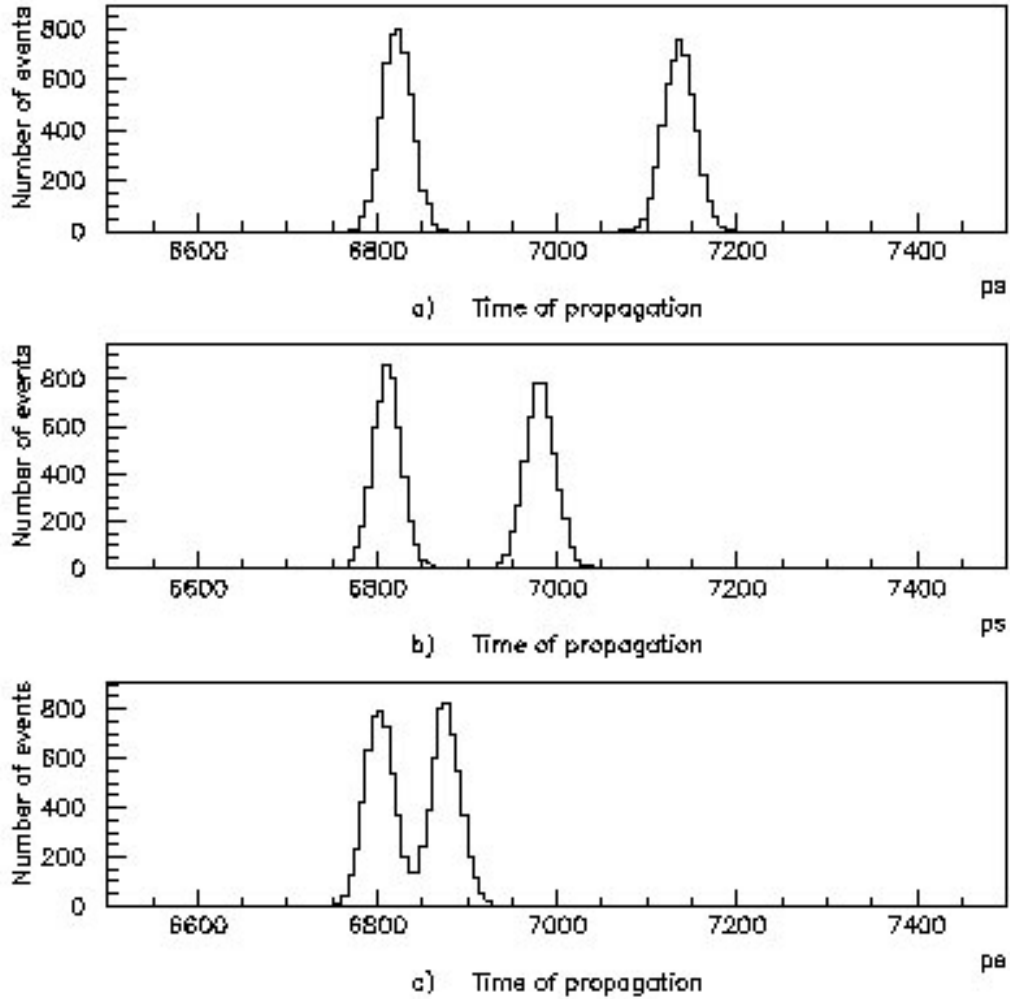

FIG. 8: Average time of propagation distributions for forward going photons with $\left|\Phi_c\right| \leq 15^\circ$ and L = 100 cm, for π (left histograms) and K (right histograms), $\theta_{inc} = 90^\circ$ and p = 1.5 (a), 2.0 (b), 3.0 (c) GeV/c momentum. Total number of events is 10000 with 50% π and 50% K tracks.

These simulations have demonstrated that such a simple and compact TOP Cherenkov counter can be used for $\pi$ and K separation in the 1-3 GeV/c momentum range. The differences of TOF between $\pi$ and K are 231 and 37 ps at 10 m time-of-flight distance and for p = 4 and 10 GeV/c, respectively, and by using 40 cm length Cherenkov detector based on the picosecond photon detector, about 25$\sigma$ and 4$\sigma$ $\pi$/K separation can be reached for p = 4 and 10 GeV/c, respectively.

## 5. Discussion and Summary

Recent R&D work has resulted in a simple and effective RF deflector, and to develop and construct the picosecond photon detector the problem is only in time and money. It is expected that the Cherenkov Correlated Timing counter or Chernkov Time Of Propagation detector with picosecond photon detector, provide not only high particle identification ability comparable with the ability of 3D DIRC, but also they are more compact and flexible detectors.

Such a counter with radiator L $\leq$ 50 cm is ideally suited as a TOF counter for PID up to a momentum of 10 GeV/c.

Although the DIRC, CCT and TOP techniques have been proposed for the $\pi$/K identification at B-factories they also can find application at fixed-target experiments. Since the incoming particles in most cases of fixed-target experiments are incident nearly normal to the detector, a maximal separation power can be obtained. CCT and TOP Cherenkov detectors can be used at CEBAF in all three Halls and in the future 12 GeV program to identify particle types. The operation of this technique with CEBAF CW photon-electron beams and with fine bunch microstructure (~2ps bunches each 2 ns period), is similar to the operation of the RF separator, if the RF deflector is operated with the same frequency as the master oscillator of the accelerator. Secondary photo-electrons produced by electrons, pions, kaons, and protons with the same momentum and flight path and from the same place of the radiator, pass the RF deflecting system are deflected by different angle and hit different places on the detector. Therefore, secondary electrons from different kinds of particles are separated in space. This will allow measurement of relative velocities, as well as organization of dedicated tagging. For example, a carefully designed spectrometer and small size Cherenkov TOF detector will respond to only one kind of particle e.g. to kaons.

AM is grateful to B. Mecking for useful discussions. This work is supported in part by International Science and Technology Center- ISTC, Project A-372.